\documentstyle[preprint,epsf,aps]{revtex}
\def\beq{\begin{equation}}
\def\eeq{\end{equation}}
\def\beqn{\begin{eqnarray}}
\def\eeqn{\end{eqnarray}}

\draft
\tighten

\begin{document}
%\begin{center}
\title{Inelastic photon scattering and the magnetic moment of 
the $\Delta$ (1232) resonance}
\author{D. Drechsel, M. Vanderhaeghen}
\address{Institut f\"{u}r Kernphysik, 
Johannes Gutenberg Universit\"{a}t, D-55099 Mainz, Germany}
\author{M.M. Giannini, E. Santopinto}
\address{Dipartimento di Fisica and Istituto Nazionale di Fisica
  Nucleare, Universit\`a di Genova, Italy}
%\end{center}
\date{\today}
\maketitle

\begin{abstract}

The reaction $\gamma+p\rightarrow\gamma'+p+\pi^0$ has been suggested  
as a means to deduce the $\Delta^+$ magnetic moment. The cross 
section for this process is estimated in both the constituent quark 
model and an effective Lagrangian procedure. The resulting total 
cross section is of the order 5-10 nb, which is at the limit of 
present experimental capabilities.

\end{abstract}

\section{Introduction}

The static properties of baryons are an important testing ground for
QCD based calculations in the confinement region. 
In particular, a recent comparison of theoretical predictions 
for $\Delta$ magnetic moments in different approaches has been given
in \cite{Ali00}. However, little experimental information is 
available for hadrons outside of the ground state SU(3) octet.   
In view of the short life-time of the resonances, such
information has to come from a detailed analysis of intermediate
states. It was therefore suggested by Kondratyuk and
Ponomarev~\cite{Kon68} to consider radiative $\pi^+p$ scattering as a
means to measure the static properties of the $\Delta$ isobar. As a
result of many experimental and theoretical efforts~\cite{Nef78}, the
Particle Data Group~\cite{PDG98} now quotes a value of
$\mu_{\Delta^{++}}= (5.6\pm 1.9)\mu_N$~\cite{PDG98} for the magnetic
dipole moment of the $\Delta^{++}$ resonance. The large error bar is
due to large nonresonant processes, external bremsstrahlung by
initial and final state particles, and a strong background due to 
interactions in both the initial and final
states~\cite{Kon68,Nef78}.  A much cleaner experiment would be an
electromagnetic excitation of the nucleon leading to the $\Delta$
resonance with subsequent emission of a real photon followed by the
decay into a nucleon and a pion. The process $\gamma+p\rightarrow
\gamma'+p'+\pi^0$ would be particularly favorable, because the signal
is less disturbed by the external bremsstrahlung background.
Machavariani et al.~\cite{Mac99} have recently investigated this
reaction in the framework of an effective Lagrangian model. For an
incident photon with lab energy $E^{lab}_{\gamma}=386$~MeV (equivalent
to a $cm$ energy $k=286$~MeV or a total $cm$ energy $W=1267$~MeV) they
find differential cross sections of the order of 0.5~nb/sr$^2$~MeV,
with a maximum for photons emitted with lab energy
$E^{lab}_{\gamma}\approx 65$~MeV and at an angle
$\Theta'\approx90^{\circ}$. If we integrate these differential cross
sections over all angles of the emitted pion and photon and over the
final state photon energy, we obtain a total cross section of several
microbarns.

Unfortunately, the A2 collaboration at MAMI working with the TAPS
detector has only been able to see 3-photon events~\cite{Met99} 
at a rate corresponding to a cross section of tens of nanobarns, partly  
originating from the reaction 
$\gamma+p\rightarrow p+\pi^0+\pi^0$~\cite{Har96}.
It is the aim of this contribution to point out that we only expect
total cross sections for this process in the range of 5-10 $nb$, which
is probably at the limit of the present experimental accuracy. For
this purpose we have calculated both elastic (Compton) and inelastic
photon scattering through the $\Delta$ resonance in the constituent
quark model~\cite{Isg78}. Though this model is somewhat naive and has
certain deficiencies, it has the advantage that the ratio of elastic
and inelastic photon scattering can be expressed analytically. As a
result we expect that at least the predicted order of magnitude of the
cross section will be realistic. We find some further support for
these results from a numerical calculation based on an effective
Lagrangian.
\section{Photon Scattering in the Constituent Quark Model}
The interaction Hamiltonian between a real photon (momentum $\vec{k}$,
polarization vector $\hat{\epsilon}_{\lambda}$) and the 3-quark system
is
\beq
{\cal H}_{\lambda}^{{\mbox{\scriptsize{int}}}} =
-\frac{1}{\sqrt{2k}}\ \hat{\varepsilon}_{\lambda} \cdot \vec{J}\ ,
\label{DD01}
\eeq
where $\vec{J}$ is the current operator,
\beq
\vec{J}=\sum_k\frac{e}{2m_k}q(k) [\vec{p}\ '(k)+\vec{p}(k)+i\vec{\sigma}(k)\times(\vec{p}\ '(k)-\vec{p}(k))]\ ,
\label{DD02}
\eeq
with $m_k$ the mass, $q(k)$ the charge in units of $e,
\vec{\sigma}(k)$ the spin operator, $\vec{p}$ and $\vec{p}\ '$ the
initial and final $cm$ momenta of quark $k$, respectively, and
$e^2/4\pi\approx1/137$ the fine structure constant. In the following
we shall perform all calculations in the $cm$ frame of the initial
photon-nucleon system. Furthermore we shall only be interested in
static properties, i.e.  magnetic moments of both ground state and
$\Delta$ resonance, and of magnetic transitions between these two
states. Neglecting then the convection currents, the Hamiltonian of
Eq.~(\ref{DD01}) can be expressed by the magnetic moment operator
$\vec{\mu}$,
\beq
{\cal H}_{\lambda}^{{\mbox{\scriptsize{int}}}} = \mp \lambda
\sqrt{\frac{k}{2}}\ \mu_{\lambda}\ ,
\label{DD03}
\eeq
with the sign corresponding to absorption and emission
respectively, and
\beq
\mu_{\lambda} = \sum_k\frac{e}{2m_k}q(k)\ \sigma_{\lambda}(k)\ .
\label{DD04}
\eeq
For simplicity we also assume that $m_k=m_N/3$, where $m_N=938$~MeV is
the mass of the proton. If we use symmetrized quark wave functions,
Eq.~(\ref{DD04}) can be cast into a form operating only on the 3rd quark,
\beq
\mu_{\lambda} = 9\ q(3)\ \sigma_{\lambda}(3)\ \mu_N\ ,
\label{DD05}
\eeq
where $\mu_N=e/2m_N$ is the nuclear magneton.

With the usual symmetrized isospin $(\Phi)$ and spin $(\chi)$ wave
functions of the 3-quark system,
\beq
\mid N \rangle = \frac{1}{\sqrt{2}}\mid \Phi_{MS} \chi_{MS} +
\Phi_{MA} \chi_{MA}\rangle\ , \mid\Delta=\mid\Phi_S\chi_S\rangle\ ,
\label{DD06}
\eeq
we obtain the well-known quark model predictions for the magnetic
moments of nucleon $(n,p)$ and Delta $(\Delta^{++},\ \Delta^{+},\
\Delta^{0},\ \Delta^{-})$,
\beqn
\mu_p & = &\langle p\frac{1}{2}\mid\mu_0\mid
p\frac{1}{2}\rangle = 3\mu_N\ ,\nonumber \\
\mu_n & = &\langle n\frac{1}{2}\mid\mu_0\mid
n\frac{1}{2}\rangle = -2\mu_N\,\nonumber \\
\mu_{\Delta^c} & = &\langle \Delta^c\frac{3}{2}\mid\mu_0\mid
\Delta^c\frac{3}{2}\rangle = 3e_c\mu_N\ ,
\label{DD07}
\eeqn
where we have indicated the spin projection by $\frac{1}{2}$ or
$\frac{3}{2}$, and $e_c$ is the charge of the respective $\Delta^c$ in
units of $e$. For further use we also evaluate the matrix elements
of $\mu_{\pm}$, in particular
\beqn
\langle \Delta^+ \frac{3}{2}\mid \mu_+\mid p \frac{1}{2} \rangle
\ \ \ & = & 2\sqrt{3}\  \mu_N\ , \nonumber \\
\langle \Delta^+ \frac{1}{2}\mid \mu_+\mid p, -\frac{1}{2} \rangle
& = & 2\  \mu_N\ .
\label{DD08}
\eeqn

The familiar helicity amplitudes are given by the matrix elements of
the Hamiltonian Eq.~(\ref{DD03}),
\beqn
A_{1/2}& = & -\sqrt{\frac{k}{2}} \langle \Delta^+
\frac{1}{2}\mid \mu_+\mid p, -\frac{1}{2} \rangle
= -\sqrt{2k}\  \mu_N\  , \nonumber \\
A_{3/2} & = &  -\sqrt{\frac{k}{2}} \langle \Delta^+
\frac{3}{2}\mid \mu_+\mid p \frac{1}{2} \rangle
= -\sqrt{6k}\  \mu_N\ .
\label{DD09}
\eeqn
The electromagnetic width for the decay of the $\Delta$ (1232) is
\beq
\Gamma_{\gamma} = \frac{k^2E}{2\pi m_{\Delta}}
\left (\mid A_{1/2} \mid^2 + \mid A_{3/2} \mid^2 \right )
=\frac{4k^3 E}{\pi m_{\Delta}} \mu_N^2  \ ,
\label{DD10}
\eeq
where $k=(m_{\Delta}^2-m_N^2)/2m_{\Delta}=259$~MeV and $E =
\sqrt{m^2_N+k^2} = 973~$MeV at resonance, $W=k+E=m_{\Delta}$.

Numerically we obtain the result $\Gamma_{\gamma}\approx
0.46$~MeV and a branching ratio $\Gamma_{\gamma}/\Gamma\approx
0.38~\%$, while the experimental value is (0.56$\pm$ 0.04)~\%,
i.e. the quark model underestimates the helicity amplitudes of
Eq.~(\ref{DD09}) by about 20~\%. If the quark mass (or and
additional quark g-factor) is adjusted to the measured magnetic
moment of the proton, the helicity amplitudes of the quark model
come out even 25~\% too low. The experimental value is obtained by
multiplying the $rhs$ of Eq.~(\ref{DD08}) by
$G_{N\Delta}/2\sqrt{3}$, with $G_{N\Delta}\approx 4$. 
We note that in calculations with a harmonic oscillator potential, the
{\it rhs} of Eqs.~(\ref{DD07})-(\ref{DD09}) is usually multiplied by a
factor $\exp ( - k^2 / 6 \alpha_0^2)$, with $\alpha_0$ the oscillator
parameter. This retardation factor adds some model dependence and
leads to a stronger underestimation of the electromagnetic decay width
$\Gamma_\gamma$, which would have to be corrected by choosing 
$G_{N\Delta}\approx 4 \exp (k^2 / 6 \alpha_0^2)$. 

Using the same matrix elements we can also evaluate the differential
cross section for forward Compton scattering,
\beqn
\frac{d\sigma^{{\mbox{\scriptsize{el}}}}}{d\Omega'dk'} (\Theta=0)&=&
\frac{m_N^2}{16\pi^2 EW} \ \frac{k'}{k} \delta (k'+E'-W) \, 
\mid G(W) \mid^2 \nonumber\\
&&\times\frac{1}{4} \sum_{\lambda M} \mid\langle pM\mid j_{-\lambda} \mid
\Delta, M+\lambda\rangle \  \langle \Delta, M+\lambda
\mid j_{\lambda} \mid \ pM\rangle \mid^2\ ,
\label{DD11}
\eeqn
with the current operator $j_{\lambda}(k)=\pm
k\lambda\mu_{\lambda}$ for absorption and emission respectively
(see Eqs.~(\ref{DD01}) to ~(\ref{DD03})). The required matrix
elements can be obtained from Eq.~(\ref{DD08}) and the symmetry
relations (valid for $\lambda=\pm1$)
\beq
\langle\Delta M' \mid \mu_{\lambda} \mid pM \rangle =
\langle\Delta, -M' \mid \mu_{-\lambda} \mid p,-M \rangle=
(-)^\lambda \langle pM \mid \mu_{-\lambda} \mid \Delta M' \rangle \ .
\label{DD12}
\eeq
The propagator in the intermediate state takes the nonrelativistic form
\beq
G(W) = (m_{\Delta}-W-\frac{i}{2} \Gamma (W))^{-1}\ ,
\label{DD13}
\eeq
with an energy-dependent width $\Gamma(W)$ fixed at
$\Gamma(m_{\Delta})=120$~MeV. For the elastic process we have $E=E'$
and $k'=k$, and using
Eqs.~(\ref{DD08},\ref{DD11},\ref{DD12},\ref{DD13}) we obtain the
following prediction for forward Compton scattering,
\beq
\frac{d\sigma^{{\mbox{\scriptsize{el}}}}}{d\Omega'} (\Theta=0) =
\sigma_{Th}\frac{5k^4}{EW[(m_{\Delta}-W)^2 + \frac{1}{4} \Gamma^2]}\ ,
\label{DD14}
\eeq
where $\sigma_{Th} = (e^2/4\pi m_N)^2\approx 23.6$~nb is the
Thomson cross section. In particular the predicted cross section
at resonance is
\beq \frac{d\sigma^{{\mbox{\scriptsize{el}}}}}{d\Omega'} (\Theta=0,
W=m_{\Delta}) \approx 125~{\mbox{nb/sr}}\ .
\label{DD15}
\eeq
With an angular distribution for purely magnetic transitions,
$(7+3\cos^2\Theta)/10$, the total Compton cross section at resonance
is $\sigma_{tot}^{el}(W=m_{\Delta})\approx 1.2~\mu$b, which has to be
compared to an experimental value of about 2.8~$\mu$b~\cite{Dre99}. The
quark model underestimates the data by a factor of about 2, because
the cross section scales with the 4th power of the helicity
amplitudes.

Along exactly the same lines we shall now estimate the inelastic
cross section for the reaction of interest. This requires the
following changes in Eq.~(\ref{DD11}):
\begin{enumerate}
\item[(a)] $\sum_{\lambda M} \rightarrow \sum_{\lambda\lambda'M}$,
\item[(b)]$\langle pM\mid j_{-\lambda}\mid \Delta, M+\lambda
\rangle \rightarrow \langle \Delta, M+\lambda -\lambda'\mid
j_{-\lambda'}\mid \Delta, M+\lambda \rangle$,
\item[(c)]$\delta(k'+E'-W) \rightarrow \frac{\Gamma
(W')}{2\pi[(m_{\Delta}-W')^2 + \frac{1}{4} \Gamma^2 (W')]}$,
\end{enumerate}
where $W'^2 = W^2 - 2 W k'$ is the total $cm$ energy of the $\pi N$ system.
With regard to (a) we note that there occur 6 ``paths'' from the
initial spin projection of the nucleon to the final spin
projection of the $\Delta$,
\beq
 \frac{1}{2}\rightarrow \frac{3}{2}\rightarrow\frac{1}{2}\ ,
-\frac{1}{2}\rightarrow +\frac{1}{2}\rightarrow +\frac{3}{2}\ ,
-\frac{1}{2}\rightarrow +\frac{1}{2}\rightarrow -\frac{1}{2}\ ,
\label{DD16}
\eeq
and 3 more that can be obtained by reversing all signs. In
addition to the matrix elements of Eq.~(\ref{DD08}) we need the
matrix elements
\beqn
\langle \Delta^+ \frac{3}{2}\mid \mu_+ \mid \Delta^+ \frac{1}{2}
\rangle& = & \sqrt{6}\mu_N \ ,\nonumber \\
\langle \Delta^+ \frac{1}{2}\mid \mu_+ \mid \Delta^+, -\frac{1}{2}
\rangle& = & 2\sqrt{2}\mu_N\ .
\label{DD17}
\eeqn
All other matrix elements follow from Eq.~(\ref{DD12}). With the above
changes the analogon of Eq.~(\ref{DD11}) for inelastic scattering is
\beqn
&&\frac{d\sigma^{{\mbox{\scriptsize{inel}}}}}{d\Omega'dk'}
(\Theta=0)= \frac{m_N^2 k k'^3}{16\pi^2 EW}\ \frac{\mid G(W)\mid^2
  \mid G(W')\mid^2 \Gamma(W')} {2\pi} \nonumber \\&& \times\frac{1}{4}
\sum_{\lambda\lambda' M} \mid\langle \Delta^+,M+\lambda-\lambda' \mid
\mu_{-\lambda'} \mid \Delta^+, M+\lambda\rangle \ \langle \Delta^+,
M+\lambda \mid \mu_{\lambda} \mid \ pM\rangle \mid^2\ .
\label{DD18}
\eeqn
By use of Eqs.~(\ref{DD08},\ref{DD17}), this expression can be cast
into the form
\beq
\frac{d\sigma^{{\mbox{\scriptsize{inel}}}}}{d\Omega'dk'} (\Theta=0)=
\frac{4 m_N^2 k k'^3}{\pi^2 EW}\ \mu_N^4
\ \mid G(W)\mid^2 \frac{\mid G(W')\mid^2 \Gamma(W')}{2\pi}\ .
\label{DD19}
\eeq
We note that in the zero-width approximation
\beq
\lim_{\Gamma\to 0} \frac{\mid G(W')\mid^2 \Gamma(W')} {2\pi}
\rightarrow \delta(m_{\Delta}-W')\ ,
\label{DD20}
\eeq
and since $k\rightarrow m_{\Delta}-E_N$ at resonance, $k'$ tends to
zero in that limit. As a result the inelastic cross section is
suppressed with regard to Compton scattering by a factor
$(k'/k)^3\approx(\frac{\Gamma}{2(m_{\Delta}-m_N)})^3\approx 10^{-2}$.

If we restrict the discussion to the $\pi^0$ channel, the density of
the final states will be further reduced by $\Gamma(\Delta^+\rightarrow
p\pi^0)/\Gamma\approx 2/3$. From a more detailed calculation
we find a rather constant angular distribution of the form
(11 -3 $\cos^2\Theta)$/8, and an integrated cross section
\beq
\frac{d\sigma^{\mbox{\scriptsize{inel}}}_{\mbox{\scriptsize{tot}}}}{dk'}
= \sigma_{Th} \frac{20 k k'^3}{3 EW}\
\frac{\Gamma(W')}{[(m_{\Delta}-W)^2+\frac{1}{4} \Gamma^2(W)] 
[(m_{\Delta}-W')^2+\frac{1}{4} \Gamma^2(W')]}\ .
\label{DD21}
\eeq

\section{Results and Discussion}

Comparing the total inelastic cross section to the total elastic one,
Eq.~(\ref{DD14}), we obtain the simple relation
\beq
R = \frac{\sigma^{{\mbox{\scriptsize{inel}}}}_{\mbox{\scriptsize{tot}}}}
{\sigma^{{\mbox{\scriptsize{el}}}}_{\mbox{\scriptsize{tot}}}
} = \frac{5}{6} \int_0^{k_{max}} dk'
\left (\frac{k'}{k}\right )^3
\frac{\Gamma(W')} {2\pi[(m_{\Delta}-W')^2
 + \frac{1}{4} \Gamma^2(W')]}\ .
\label{DD22}
\eeq
The upper limit of the integral is given by the threshold of $\pi^0$
production, $W'=m_N+m_{\pi^0}$, at which point the energy-dependent
width vanishes with the 3rd power of the pion momentum.  The maximum
inelastic cross section is expected for $W\approx
m_{\Delta}+\Gamma/2$, i.e. $k\approx 306$~MeV and $k'_{max} \approx
200$~MeV. If we increase the initial photon energy, the phase space of
Eq.~(\ref{DD19}) increases, but the excitation strength decreases,
because we move away from the resonance. The opposite is true if the
initial photon energy is decreased, the increase in excitation
strength is compensated by a shrinking of the phase space.  At
energies $k\approx 300$~MeV, the ratio of Eq.~(\ref{DD22}) has a
value of about $6.4 \cdot 10^{-3}$ if one uses a typical energy-dependent
width.

We recall at this point that the quark model predictions for the
elastic and inelastic cross sections should be scaled by
$(G_{N\Delta}/2\sqrt{3})^4$ and $(G_{N\Delta}/2\sqrt{3})^2
(\mu_{\Delta^+}/\mu_p)^2$ respectively, in order to describe the
experimentally observed cross section. 
As a result we therefore predict a total inelastic cross section
\beq
\sigma^{{\mbox{\scriptsize{inel}}}}_{\mbox{\scriptsize{tot}}}\approx
R\frac{(\mu_{\Delta^+}/\mu_p)^2}{(G_{N\Delta}/2\sqrt{3})^2}\
\sigma^{{\mbox{\scriptsize{el}}}}_{\mbox{\scriptsize{tot}}}\approx 4
\left (\frac{\mu_{\Delta^+}}{\mu_p} \right )^2\ nb\ ,
\label{DD23}
\eeq
with
$\sigma^{{\mbox{\scriptsize{el}}}}_{\mbox{\scriptsize{tot}}}\approx
0.9~\mu$b the Compton cross section at $k\approx 300$~MeV.

As a further check we have evaluated the cross section for the
reaction $\gamma p\rightarrow\Delta^+\rightarrow\gamma\Delta^+
\rightarrow \gamma \pi^0 p$ in
the framework of an effective Lagrangian (see e.g. \cite{Van95} for
details). For the $\Delta$ current we take the simple form
\beq
\langle \Delta^+(p',s')\mid J^{\mu}(0)\mid \Delta^+(p,s)\rangle
= +\frac{e}{2m_{\Delta}} \bar{\Delta}_{\alpha} (p',s')
\{ (p'+p)^{\mu} + i\sigma^{\mu\nu} (p'-p)_{\nu} G_{M1}^{\Delta} \}
\Delta^{\alpha} (p,s)\ ,
\label{DD24}
\eeq
where $G_{M1}^{\Delta}$ is the magnetic moment of the $\Delta^+$ in
units of the ``$\Delta$ magneton'', $e/2m_{\Delta}$. 

\begin{figure}[ht]
\epsfxsize=14 cm
\centerline{\epsffile{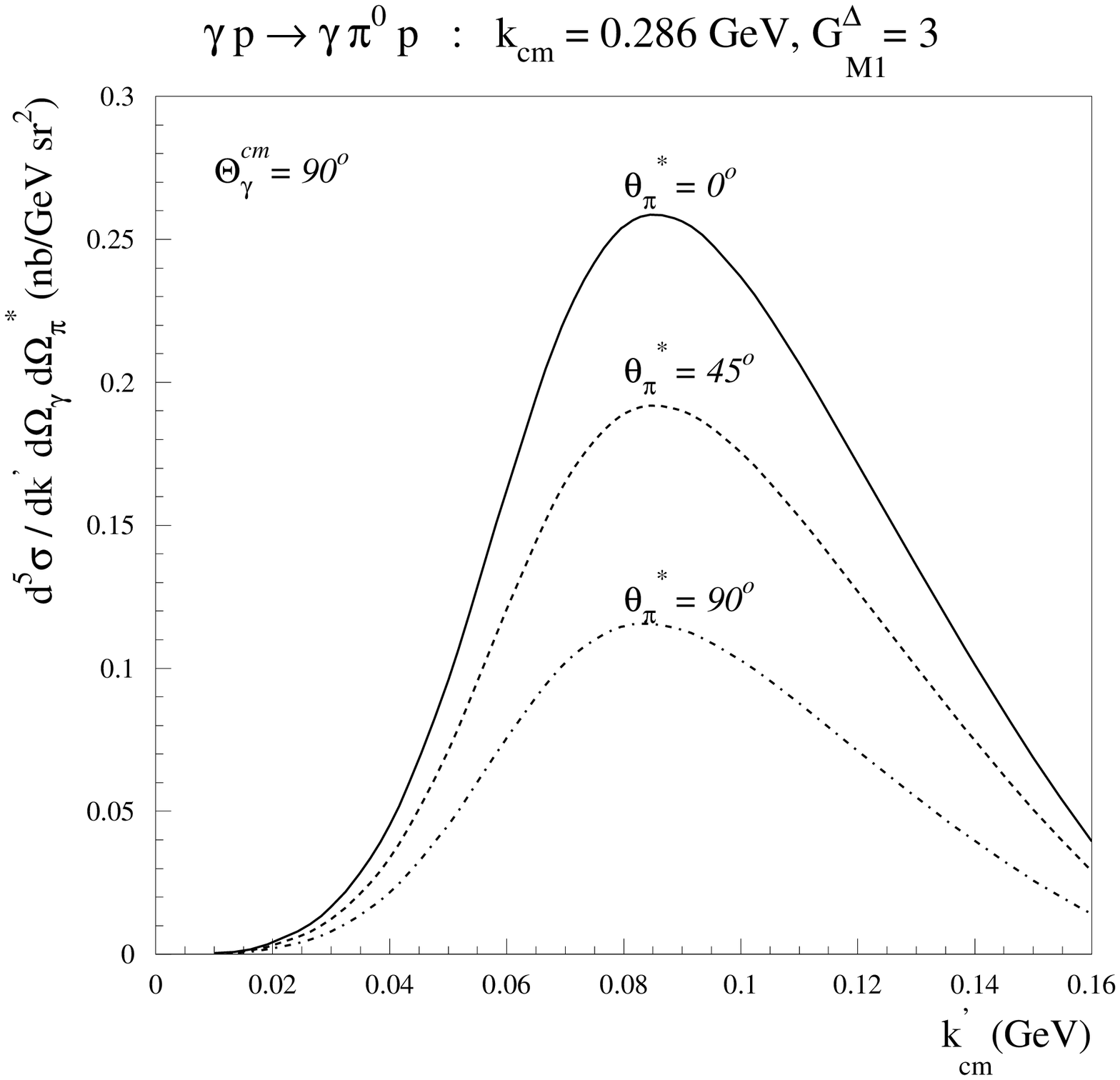}}
\vspace{-1.cm}
\caption[]{Photon energy dependence of the 
fivefold $\gamma p \to \gamma \Delta^+ \to \gamma \pi^0 p$ 
differential cross section with outgoing photon energy and angle in $\gamma p$
  {\it cm} system, and with the pion angles in the $\pi^0 p$ rest frame. }
\label{fig:spectral}
\end{figure}

\begin{figure}[ht]
\epsfxsize=14 cm
\centerline{\epsffile{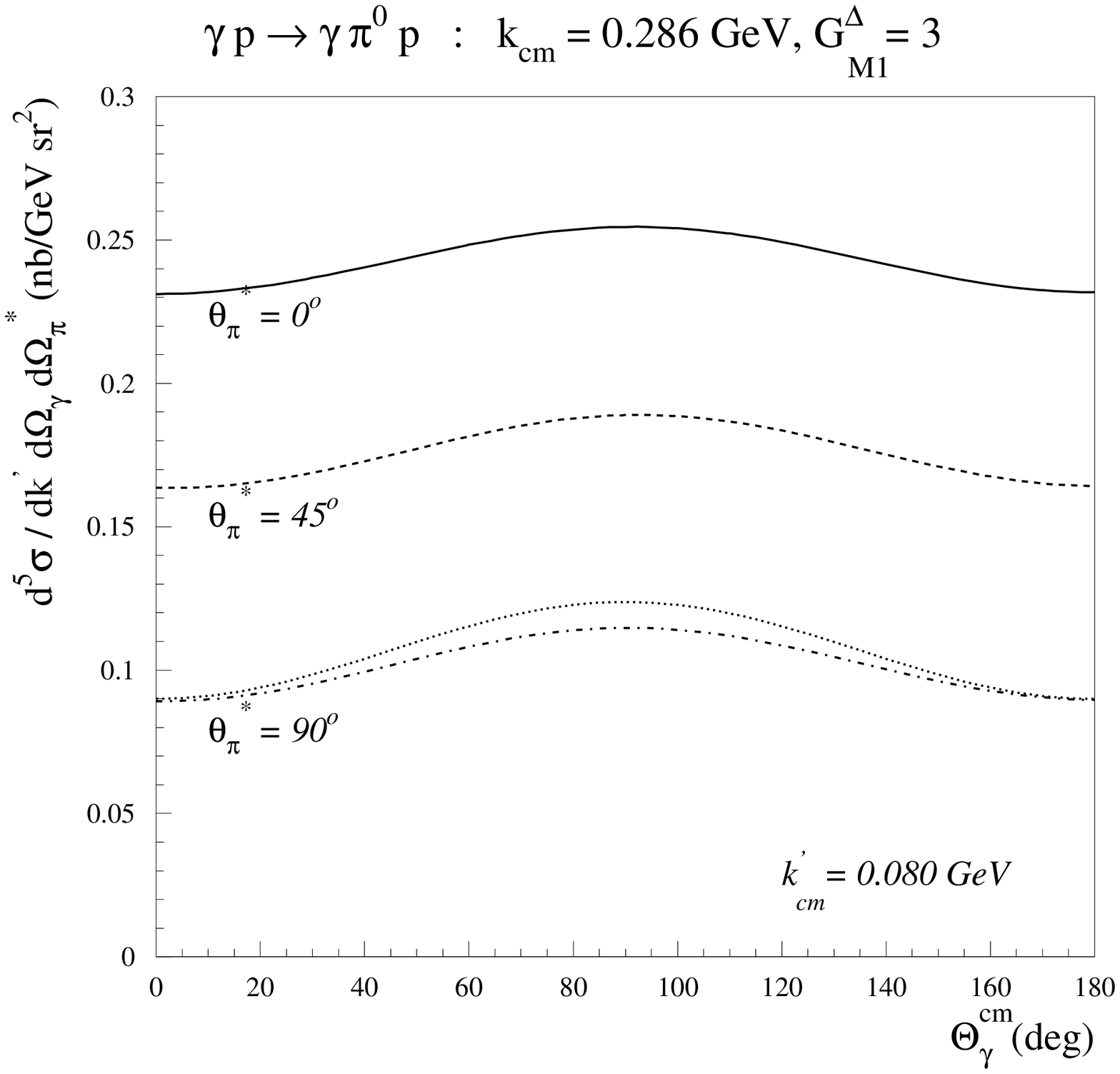}}
\vspace{-1.cm}
\caption[]{Photon angular dependence of the 
fivefold $\gamma p \to \gamma \Delta^+ \to \gamma \pi^0 p$ 
differential cross section. For comparison, the
angular dependence $(11 - 3 \cos^2\Theta)/8$ in the quark model
  calculation is shown by the dotted curve (with a
global rescaling factor to match the lower curve).}
\label{fig:angular}
\end{figure}

In Figs.~\ref{fig:spectral}-\ref{fig:angular}, the differential 
cross sections are shown following from the effective Lagrangian for
the value $G_{M1}^{\Delta} = 3$.

In Fig.~\ref{fig:spectral}, we show the fivefold differential cross
section at an incoming photon {\it cm} energy of $k$ = 286 MeV (as
also considered in \cite{Mac99}), at a photon {\it cm} angle of 90$^o$
and for three in-plane pion angles (in the $\pi^0 p$ rest frame).  The
energy dependence of the cross section reflects the invariant mass
distribution of the $\pi^0 p$ system and has a maximum around an
outgoing photon $cm$ energy of 80 MeV. While this energy dependence is
in good agreement with the results of Ref.~\cite{Mac99}, we obtain an
absolute value for the cross section that is lower by about 3 orders
of magnitude. At this energy, we show in Fig.~\ref{fig:angular} the
photon angular dependence, which is seen to be quite flat with a
slight maximum at 90$^o$. For comparison, the angular dependence is
well approximated in the quark model calculation by its form $(22 - 6
\cos^2\Theta)/16$, which is shown on the same figure (dotted curve,
with a global rescaling factor to match the lower curve in
Fig.~\ref{fig:angular}). Comparing with Ref.~\cite{Mac99}, our
cross sections are again lower by nearly 3 orders of magnitude.
Moreover the angular distribution given in Ref.~\cite{Mac99} has a
high maximum near 90$^{\circ}$ (in the {\it lab}) 
and approaches very small values in the forward and backward directions, 
quite at variance with our nearly constant angular distributions.

\begin{figure}[ht]
\epsfxsize=14 cm
\centerline{\epsffile{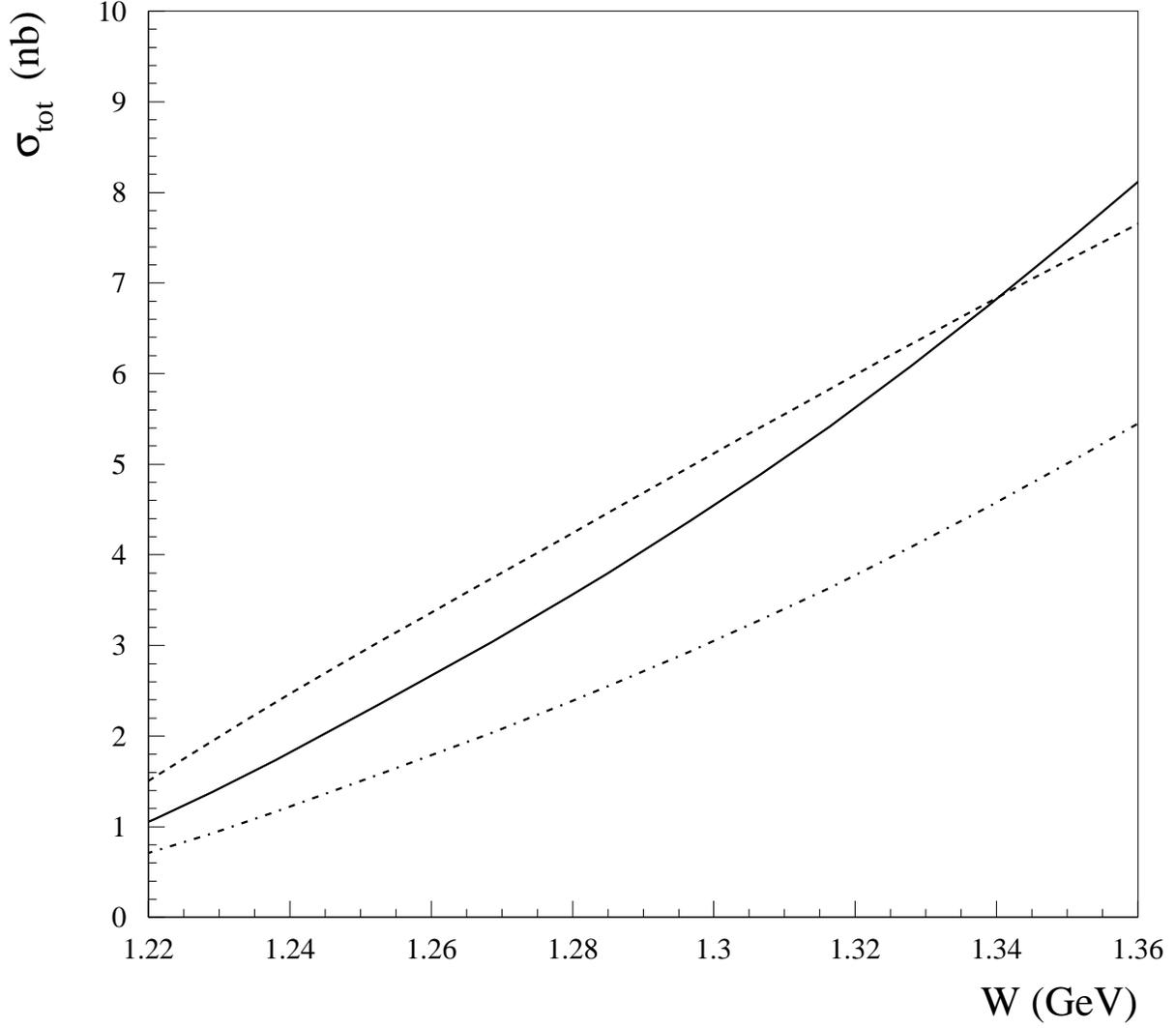}}
\vspace{-1.cm}
\caption[]{Total cross section for the 
$\gamma p \to \gamma \Delta^+ \to \gamma \pi^0 p$ reaction as function 
of the total {\it cm} energy $W$. 
The quark model calculation (with $\mu_{\Delta^+} =
\mu_p$) is shown by the dashed curve. 
The effective Lagrangian calculation is shown both for 
$G_{M1}^{\Delta} = 3$ (dashed-dotted curve) and 
$G_{M1}^{\Delta} = 3.66$ 
(corresponding with $\mu_{\Delta^+} = \mu_p$, full curve).}
\label{fig:totcross}
\end{figure}

In Fig.~\ref{fig:totcross}, we compare the quark model prediction of
Eq.~(\ref{DD23}) for the total 
$\gamma p\rightarrow 
\gamma \pi^0 p$ cross section (using $\mu_{\Delta^+} = \mu_p$)
with the result from the effective Lagrangian calculation as function
of the total {\it cm} energy $W$. 
The effective Lagrangian calculation is shown both for a value 
$G_{M1}^{\Delta} = 3$ and 
$G_{M1}^{\Delta} = 3.66$ (which corresponds with 
$\mu_{\Delta^+} = \mu_p$). 
One sees that both calculations are in rather good agreement and yield
total cross sections of the order 5~nb in the considered energy
range when using $\mu_{\Delta^+} = \mu_p$).

\section{Conclusions}
In spite of new and improved experimental techniques, the reaction
$\gamma+p\rightarrow\gamma'+\Delta^+\rightarrow\gamma'+p+\pi^0$ has
not yet been observed. Our expectations based on a simple constituent
quark model show that the integrated cross section for this process
should indeed be very small, namely of the order of 5~nb. These result
are corroborated by an equally simple Lagrangian approach, which also
permits us to calculate the five-fold differential cross sections with
the result of typically 0.25~nb/GeV sr$^2$. In agreement with the quark
model prediction, the angular distribution for photon emission is
rather constant.

In the future we plan to improve our calculations by including the
backgrounds due to bremsstrahlung off the incoming and outgoing
protons as well as radiation from nonresonant intermediate states.
Though we do not expect qualitative changes by such terms, they are
likely to influence the angular and energy distributions of the
process. A better understanding of these backgrounds might also give a
chance to analyze the data obtained with incident photons of higher
energies, which would increase the counting rate by giving a larger
phase space to the emitted particles.

We conclude that the electromagnetic moments of baryon resonances are
among the most evasive properties of hadrons. The extremely weak
signals for these moments are at the very limits of even the most
advanced experimental techniques. However, such data would be
invaluable for our understanding of QCD in the confinement region, and
therefore a dedicated experiment is certainly desirable.

\section*{Acknowledgements}

It is a pleasure to acknowledge the fruitful conversations with
members of the A2 Collaboration at MAMI, in particular for private
communications given by M. Kotulla and V. Metag. 
D.D. thanks the INFN and physics
department of Genova University for their hospitality. 
This work was supported in part 
by the Deutsche Forschungsgemeinschaft (SFB 443) and the 
Istituto Nazionale di Fisica Nucleare (INFN).

\end{document}